\documentclass[11pt]{article}
\usepackage{amsfonts,amsmath}
\usepackage[vcentermath]{youngtab}
\usepackage{young}
\def\bbt{\bibitem}
\def\be{\begin{equation}}
\def\en{\end{equation}}
\def\ber{\begin{eqnarray}}
\def\enr{\end{eqnarray}}
\def\nmb{ \nonumber\\}
\def\d{\partial}

\def\rbrc{\rbrace}
\def\lbrc{\lbrace}
\def\ov{\over }

\def\sgm{\sigma}
\def\Sgm{\Sigma}
\def\al{\alpha}
\def\bet{\beta}
\def\gm{\gamma}

\def\lm{\lambda}
\def\Lm{\Lambda}
\def\Om{\Omega}
\def\om{\omega}
\def\et{\eta}

\def\Tt{\Theta}

\def\dlt{\delta}

\begin{document}
\vskip 2 true cm

\centerline{\bf Line bundle twisted chiral de Rham complex}
\centerline{\bf and bound states of D-branes on toric manifolds.}

\vskip 1.5 true cm
\centerline{\bf S. E. Parkhomenko}
\centerline{\it Landau Institute for Theoretical Physics, 142432 Chernogolovka of Moscow region, Russia}
\centerline{\it Moscow Institute of Physics and Technology, 141707 Dolgoprudny of Moscow Region, Russia}
\vskip 0.5 true cm
\centerline{spark@itp.ac.ru}
\vskip 1 true cm
\centerline{\bf Abstract}
\vskip 0.5 true cm

 In this note we calculate elliptic genus in various examples of twisted chiral de Rham complex on two dimensional toric compact manifolds and Calabi-Yau hypersurfaces in toric manifolds. At first the elliptic genus is calculated for the line bundle twisted chiral de Rham complex on a compact smooth toric manifold and $K3$ hypersurface in $\mathbb{P}^{3}$. Then we twist chiral de Rham complex by sheaves localized on positive codimension submanifolds in $\mathbb{P}^{2}$ and calculate in each case the elliptic genus. In the last example the elliptic genus of chiral de Rham complex on $\mathbb{P}^{2}$ twisted by $SL(N)$ vector bundle with instanton number $k$ is calculated.  In all cases considered we find the infinite tower of open string oscillator contributions and identify directly the open string boundary conditions of the corresponding bound state of $D$-branes.


\vskip 10pt

"{\it PACS: 11.25Hf; 11.25 Pm.}"


\smallskip
\vskip 10pt
\centerline{\bf 1. Introduction}
\vskip 10pt

 It has been proposed by J.Harvey and G.Moore \cite{HM} that sheaves can be used to model $D$-branes on large-radius Calabi-Yau manifolds. Since then, a significant progress has been made in understanding of sheaves as models of $D$-branes. As a review of the results and a source of necessary references see \cite{Asp}, \cite{Sh}. However, the direct systematic map between open string boundary conditions and sheaves is not known auntil now. 
 
 In the important work of Malikov, Schechtman and Vaintrob \cite{MSV} a sheaf of vertex algebras, which is called chiral de Rham complex has been introduced for every smooth variety. When the variety is $\mathbb{C}^{d}$ this sheaf is known as $"bc\bet\gm"$-system.
Soon after the significant application of chiral de Rham complex in the String Theory has been represented in the beautiful paper of Borisov \cite{B1} where the chiral de Rham complex construction has been given for each pair of dual reflexive polytopes
defining toric CY manifold. Thus Borisov constructed directly holomorphic sector of the CFT from toric data of CY manifold. 

 Another application of chiral de Rham complex appears in the problem of Gepner models
\cite{Gep} geometry investigation. The significant step in this direction has been made in the paper~\cite{GorbM} where the vertex algebra of certain Landau-Ginzburg orbifold has been related to the cohomology of the chiral de Rham complex of toric CY manifold by a spectral sequence. One of the key points of ~\cite{GorbM} is that the free-field representation of the corresponding Landau-Ginzburg orbifold is given by a number of $N=2$ minimal model  $"bc\bet\gm"$ representations of \cite{FeS}. Later \cite{P1}, $"bc\bet\gm"$ representations of $N=2$ minimal models and chiral de Rham complex have been used to investigate geometry for more complicated class of Gepner models. One should emphasize here that the results of \cite{GorbM} and \cite{P1} dealth with only holomorphic (or anti-holomorphic) sector of the $\sgm$-model and the problem how to combine chiral de Rham complex descriptions for holomorphic and anti-holomorphic sectors to get a full $\sgm$-model space of states description as well as correlation functions is still unsolved and intersting. This important problem  has been investigated in the works of Frenkel, Losev and Nekrasov \cite{FLN1}, \cite{FLN2}.
In a more general string theory context, $"bc\bet\gm"$ systems have been discussed in \cite{N}.

 In the paper \cite{P2} the generalization of Borisov construction \cite{B1} has been represented to include chiral de Rham complex on toric manifold twisted by line bundle. It was conjectured there that the cohomology of line bundle twisted chiral de Rham complex may describe an infinite tower of states in the open string sector of certain $D$-brane bound state on toric manifold. In that sense, the conjecture from \cite{P2} is an extended version of the suggestion of J.A.Harvey and G.Moore \cite{HM} allowing probably to establish the above mentioned map between sheaves and open string boundary conditions in a more systematic way. In defense of the conjecture the results of the paper \cite{P3} talk also. In that paper the open string sector of Gepner model boundary states \cite{ReS} was investigated by $"bc\bet\gm"$ representations of $N=2$ minimal models. That was possible to do because for each pair of Gepner model boundary states the open string sector was given by certain combination of (GSO projected) tensor products of N=2 minimal models representations realized by $"bc\bet\gm"$ fields due to \cite{FeS}. It allowed to show in particular \cite{P3} that the open string sector can be described as a representation of the cohomology of the chiral de Rham complex on the Landau-Ginzburg orbifold which is related to the cohomology of the chiral de Rham complex of the corresponding toric CY manifold by a spectral sequence from \cite{GorbM}. So the conjucture from \cite{P2} implies that if we fix a pair of Gepner model boundary states then the analog of spectral sequence of \cite{GorbM} which relates the corresponding open string sector to the cohomology of the chiral de Rham complex twisted by certain Chan-Paton bundle or sheaf should exist in certain large radius limit of boundary sigma model on toric manifold. The proof (or disproof) of this conjecture deserves special investigation by the ideas and methods developed in \cite{FLN1}, \cite{FLN2} but it is beyond the scope of the present paper.

 In this note we represent some additional evidences in support of the conjecture from \cite{P2} calculating elliptic genus of the chiral de Rham complex twisted by line bundle as well as more general sheaves determined on compact toric manifold and Calabi-Yau hypersurface embedded in toric manifold. In all cases considered we interprate the results 
in terms of open string oscillator contributions coming from certain bound states of $D$-branes establishing thereby a correspondence between the boundary conditions and Chern classes of Chan-Paton bundles or sheaves. Thus if we replace the usual bundles or sheaves by twisted chiral de Rham complex we include all tower of string exitations and recover the open string boundary conditions.

 We begin in section 2 with an overview of general elliptic genus formula for the line bundle twisted chiral de Rham complex on toric manifold and CY hypersurface obtained in \cite{P2} and represent some evidences why the open string states on toric manifold with holomorphic Chan-Paton bundle can be described locally by twisted chiral de Rham complex. In section 3 we calculate the elliptic genus of line bundle twisted chiral de Rham complex on $\mathbb{P}^{2}$ and generalize the result for an arbitrary compact smooth two dimensional toric variety. The result of calculations is represented in terms of infinite tower of open string oscillator contributions coming from the bound state of $D0$-$D2$-$D4$-branes. Then the elliptic genus calculation is made for the line bundle twisted chiral de Rham complex on $K3$ hypersurface embedded in $\mathbb{P}^{3}$. In this case we extract the corresponding open string oscillator contributions coming from bound state of $D$-branes also. In section 4 we generalize the results of section 3 to include more general examples of twisting sheaves. The explicit elliptic genus calculations are made in three examples. In the first example the twisting sheaf localized on a curve in $\mathbb{P}^{2}$ and we find the open string contributions from $D0$-$D2$ bound state. In the second example the twisting sheaf localized on points in $\mathbb{P}^{2}$ so the open string oscillator contributions come from bound state of $D0$-branes. In the third example the chiral de Rham complex is twisted by a sheaf of $SL(N)$ vector bundle with the instanton number $k$. As a result of the elliptic genus calculation we find the infinite tower of open string oscillator contributions coming from the bound states of $k$ D0-branes and $N$ $D4$-branes which is in agreement with the conjecture of Witten \cite{W} on the relation between the instantons and $D$-branes. Tachyon condensation picture is discussed briefly at the end of the section. We conclude in section 5.

\vskip 10pt
\centerline{\bf 2. Line bundle twisted chiral de Rham complex elliptic genus.}
\vskip 10pt

 In this section a brief review of line bundle twisted chiral de Rham complex construction and elliptic genus calculation is represented for a smooth complete toric variety. For more details the reader is referred to \cite{P2}, \cite{MSV}, \cite{B1}, \cite{B2}.

\vskip 10pt
\leftline{\it 2.1. The elliptic genus of chiral de Rham complex.} 
\vskip 10pt

 We describe first the chiral de Rham complex and elliptic genus for the complete smooth toric manifold following closely to \cite{B1}, \cite{B2}.
  
 Let $X$ be a smooth variety of dimension $d$. In local coordinates $x_{1},...,x_{d}$
the set of local sections of chiral de Rham complex on $X$, $MSV(X)$ can be described as follows. To the coordinates $x_{1},...,x_{d}$ we associate $"bc\bet\gm"$ system of fields
\ber
a_{\mu}(z)=\sum_{n}a_{\mu}[n]z^{-n},
a^{*}_{\mu}(z)=\sum_{n}a^{*}_{\mu}[n]z^{-n-1},
\nmb
\al_{\mu}(z)=\sum_{n}\al_{\mu}[n]z^{-n-{1\ov 2}}, \
\al^{*}_{\mu}(z)=\sum_{n}\al_{\mu}[n]z^{-n-{1\ov 2}},
\label{2.bcbtgm}
\enr 
$\mu=1,...,d$.
with the following nontrivial super-commutators between the modes
\ber
\lbrack a^{*}_{\mu}[n],a_{\nu}[m]\rbrack _{-}=\dlt_{\mu,\nu}\dlt(n+m)
\nmb
\lbrack \al^{*}_{\mu}[n],\al_{\nu}[m]\rbrack _{+}=\dlt_{\mu,\nu}\dlt(n+m)
\label{2.commut}
\enr
Then the set of local sections $M$ of the chiral de Rham complex is generated by the creation
operators of the fields (\ref{2.bcbtgm}) from the vacuum state $|0>$ which is defined by
\ber
a_{\mu}[n]|0>=a^{*}_{\mu}[n-1]|0>=\al_{\mu}[n]|0>=\al^{*}_{\mu}[n-1]|0>=0,
\ n>0.
\label{2.Fock}
\enr
 The important property is the behaviour of the $bc\bet\gm$ system under the local change of coordinates \cite{MSV}. For each new set of coordinates
\ber
y_{\mu}=g_{\mu}(x_{1},...,x_{d}), \
x_{\mu}=f_{\mu}(y_{1},...,y_{d})
\label{2.coordtr}
\enr
the isomorphic $bc\bet\gm$ system of fields is given by
\ber
b_{\mu}(z)=g_{\mu}(a_{1}(z),...,a_{d}(z)),
\nmb
\bet_{\mu}(z)={\d g_{\mu}\ov \d a_{\nu}}(a_{1}(z),...,a_{d}(z))\al_{\nu}(z), \
\bet^{*}_{\mu}(z)={\d f_{\nu}\ov \d b_{\mu}}(a_{1}(z),...,a_{d}(z))\al^{*}_{\nu}(z),
\nmb
b^{*}_{\mu}(z)={\d f_{\nu}\ov \d b_{\mu}}(a_{1}(z),...,a_{d}(z))a^{*}_{\nu}(z)+
{\d^{2} f_{\lm}\ov\d b_{\mu}\d b_{\nu}}{\d g_{\nu}\ov\d a_{\rho}}(a_{1}(z),...,a_{d}(z))\al^{*}_{\lm}(z)\al_{\rho}(z)
\label{2.coordtr1}
\enr

 On the space of local sections of $MSV(X)$ the N=2 Virasoro superalgebra is acting by
\ber
G^{-}=\sum_{\mu}\al_{\mu}a^{*}_{\mu}, \
G^{+}=-\sum_{\mu}\al^{*}_{\mu}\d a_{\mu}, 
\
J=\sum_{\mu}\al^{*}_{\mu}\al_{\mu},
\nmb
T=\sum_{\mu}(a^{*}_{\mu}\d a_{\mu}+\frac{1}{2}(\d\al^{*}_{\mu}\al_{\mu}-
\al^{*}_{\mu}\d\al_{\mu}))
\label{2.btgmvir}
\enr
Though this algebra is globally defined only for Calabi-Yau manifold \cite{MSV}, \cite{B1}
the zero mode $L[0]$ of the field $T(z)$ and zero mode $J[0]$ of the field $J(z)$ are invariant under the local change of coordinates and hence, globally defined in general case. It provides
the space of local sections of the chiral de Rham complex with the double grading. The $L[0]=0$ part is isomorphic to the usual de Rham complex with differential $G^{-}[0]$. Due to this double grading chiral de Rham complex possesses a natural filtration such that the graded object isomorphic to the sheaf
\ber
E(q,y)=y^{-\frac{d}{2}}\otimes_{n\geq 1}\wedge (yq^{n-1}T^{*})
\otimes_{n\geq 1}\wedge (y^{-1}q^{n}T)\otimes_{n\geq 1}Sym (q^{n}T^{*})\otimes_{n\geq 1}Sym (q^{n}T)
\label{2.elldf1}
\enr
where the powers of $y$ and $q$ are given by eigenvalues of $J[0]$ and $L[0]$ \cite{MSV}. 

 Since the cohomology $H^{*}(MSV(X))$ of chiral de Rham complex are finite-dimensional vector spaces at every eigenvalue of $L[0]$ the Euler characteristics of the chiral de Rham complex is well-defined and coincides with the Euler characteristics of the sheaf (\ref{2.elldf1}). It allows to give the following definition of the elliptic genus of the chiral de Rham complex \cite{B2}
\ber
Ell(X,y,q)=y^{-{d\ov 2}}SuperTr_{H^{*}(MSV(X))}(y^{J[0]}q^{L[0]})
\label{2.elldf2}
\enr
 
 When $X$ is a complete toric variety \cite{Dan}, \cite{Ful} the cohomology $H^{*}(MSV(X))$ could be calculated as $\check{C}$hech cohomology for the open affine covering defined by $d$-dimensional cones \cite{B2}. So one has to describe the toric data defining $X$ as well as its covering (see \cite{Dan}, \cite{Ful}). 

 We have a lattice $\Lm$ of rank $d$, its dual lattice $\Lm^{*}$ and a complete polyhedral fan $\Sgm \subset\Lm$ which is a union of finite number of $d$-dimensional cones 
\ber
\Sgm=\cup_{I} C_{I}
\label{2.Fan}
\enr
so that each intersection of the cones $C_{I}$ is also a cone from $\Sgm$.
The variety $X$ is smooth if the cones $C_{I}$ are simplicial and are generated by a basis in
$\Lm$. The cones $C_{I}$ define the open affine covering of $X$
\ber
X=\cup_{I}A_{I}, 
\ 
A_{I}=Spec(\mathbb{C}[C^{*}_{I}])
\label{2.cover}
\enr
where $C^{*}_{I}\subset \Lm^{*}$ is dual cone to $C_{I}$ \cite{Dan}, \cite{Ful}. Intersection of any number of $A_{I}$
is another open subset of this type so the covering is acyclic for $MSV(X)$ \cite{B1}. 
In addition the natural action of $(\mathbb{C}^{*})^{d}$ can be extended to $MSV(X)$. For any
affine subset $A_{I}$ this action endows the sections of $MSV(X)$ over $A_{I}$ with natural grading by the lattice $\Lm^{*}$. The same is true for the sections over an intersection of finite number of $A_{I}$'s. Thus we come to the expression from \cite{B2}
\ber
Ell(X,y,q)=y^{-{d\ov 2}}\sum_{p^{*}\in\Lm^{*}}\sum_{I_{1},...,I_{k}}(-1)^{k}sdim_{p^{*}}H^{0}(A_{I_{1}}\cap...\cap A_{I_{k}}, MSV(X))
\label{2.ellChech}
\enr
where sdim is a super-dimension which is finite for each fixed $p^{*}\in\Lm^{*}$ and powers
of $y$ and $q$. One can in fact to calculate the sum for all $p^{*}$ simultaneously by introducing a multi-variable t
\ber
Ell(X,t,y,q)=y^{-{d\ov 2}}\sum_{p^{*}\in\Lm^{*}}\sum_{I_{1},...,I_{k}}(-1)^{k}t^{p^{*}}sdim_{p^{*}}H^{0}(A_{I_{1}}\cap...\cap A_{I_{k}}, MSV(X))
\label{2.ellChecht}
\enr
and putting in the end of calculation $t=1$ \cite{B2}.

 As an important illustration of the calculation we consider $A_{I}$. Because of $A_{I}=Spec(\mathbb{C}[C^{*}_{I}])$ it is isomorphic to the affine space
with coordinates $x_{1},...,x_{d}$ so that the set of sections $M_{I}$ of $MSV(X)$ over this space has already been described by (\ref{2.bcbtgm}), (\ref{2.commut}) and (\ref{2.Fock}).
If $p_{1}^{*},...,p_{d}^{*}$ is some basis generating the cone $C^{*}_{I}$ then
\ber
\sum_{p^{*}\in\Lm^{*}}t^{p^{*}}sdim_{p^{*}}H^{0}(A_{I},MSV(X))=
\prod_{\mu=1,...,d}\prod_{k\geq 1}
\frac{(1-t^{p^{*}_{\mu}}yq^{k-1})}
{(1-t^{p^{*}_{\mu}}q^{k-1})}
\frac{(1-t^{-p^{*}_{\mu}}y^{-1}q^{k})}
{(1-t^{-p^{*}_{\mu}}q^{k})}
\label{2.ellAI}
\enr

 To include positive codimension cones contribution we use "truly remarkable identity"
\cite{KacW} (see also \cite{B2})
\ber
\prod_{k\geq 1}\frac{(1-tyq^{k-1})}{(1-tq^{k-1})}
\frac{(1-t^{-1}y^{-1}q^{k})}
{(1-t^{-1}q^{k})}=
\sum_{n\in\mathbb{Z}}t^{n}(1-yq^{n})^{-1}
\prod_{k\geq 1}
\frac{(1-yq^{k-1})(1-y^{-1}q^{k})}
{(1-q^{k})^{2}}
\label{2.Ident}
\enr 
 
 Thus we obtain for any cone $C\subset \Sgm$ \cite{B2}
\ber
\sum_{p^{*}\in\Lm^{*}}t^{p^{*}}sdim_{p^{*}}H^{0}(A_{C},MSV(X))=
\sum_{p^{*}\in\Lm^{*}}t^{p^{*}}\prod_{\mu=1,...,dimC}\frac{1}{1-yq^{p^{*}(e_{\mu})}}G(y,q)^{d}
\label{2.ellC}
\enr
where
\ber
G(y,q)=\prod_{k\geq 1}\frac{(1-yq^{k-1})(1-y^{-1}q^{k})}{(1-q^{k})^{2}}
\label{2.G}
\enr
and $e_{i}$ are generators of the cone $C$.

 Collecting the all cones contributions we find \cite{B2}
\ber
Ell(X,t,y,q)=
\nmb
y^{-d\ov 2}\sum_{p^{*}\in\Lm^{*}}\sum_{C\subset \Sgm}(-1)^{codimC}\prod_{\mu=1,...,dimC}\frac{t^{p^{*}}}{1-yq^{p^{*}(e_{\mu})}}G(y,q)^{d}
\label{2.ellt}
\enr

\vskip 10pt
\leftline{\it 2.2. The elliptic genus of line bundle twisted chiral de Rham complex.} 
\vskip 10pt

 The line bundle on a toric variety is given by toric divisor support function $\om^{*}$
\cite{Dan}, \cite{Ful}. It is a piece-wise linear function on maximal dimension cones which is consistent on the intersections of the cones. In other words the function $\om^{*}$ is a collection of elements $\lbrc \om^{*}_{I} \rbrc $ from the lattice $\Lm^{*}$ which are compatible with the restriction map for every intersection $C_{I}\cap C_{J}$:
\ber
\om^{*}_{I}|_{C_{J}}=\om^{*}_{J}|_{C_{I}}\equiv \om^{*}_{IJ}
\label{2.supportf}
\enr

 The generalization of (\ref{2.ellt}) for the line bundle twisted chiral de Rham complex
is very simple and given by \cite{P2}
\ber
Ell_{\om^{*}}(X,t,y,q)=
y^{-d\ov 2}\sum_{p^{*}\in\Lm^{*}}\sum_{C\subset \Sgm}(-1)^{codimC}\prod_{\mu=1,...,dimC}\frac{t^{p^{*}-\om^{*}_{C}}}{1-yq^{p^{*}(e_{\mu})}}G(y,q)^{d}
\label{2.twistellt}
\enr 
where $\om^{*}_{C}$ is the restriction of $\om^{*}$ on the cone $C$.

 To explain this formula we consider first the sections of line bundle twisted chiral de Rham complex over the $A_{I}$. In this case the vacuum state is 
\ber
|\Om_{I}>=\prod a_{I\mu}[0]^{-\om^{*}_{I}(e_{\mu})}|0>
\label{2.vacc}
\enr
To generate the sections one has to apply the creation operators of the $"bc\bet\gm"$ (\ref{2.bcbtgm}) to vacuum $|\Om_{I}>$ where instead of the fields $a^{*}_{\mu}(z)$ one has to take covariant derivatives fields
\ber
\nabla _{I\mu}(z)=a^{*}_{I\mu}(z)+\om^{*}_{I}(e_{\mu})z^{-1}a^{-1}_{I\mu}(z)
\label{2.nabla}
\enr
The last term in this expression is caused by a gauge potential defined on $A_{I}$.
Let us denote the module generated by this way as $\mathbb{M}_{I}$. One can show that the vacuum $|\Om_{I}>\in \mathbb{M}_{I}$ defines trivializing isomorphism of modules (over the chiral de Rham complex on $A_{C_{I}}$) \cite{P2}
\ber
g_{I}: \mathbb{M}_{I}\rightarrow M_{I}
\label{2.trivmap}
\enr
by the rule
\ber
g_{I}|\Om_{I}>=|0>,
\nmb
g_{I}(\nabla_{I\mu}[k])g_{I}^{-1}=a^{*}_{I\mu}[k], \
g_{I}(a_{I\mu}[k])g_{I}^{-1}=a_{I\mu}[k], 
\nmb
g_{I}(\al_{I\mu}[k])g_{I}^{-1}=\al_{I\mu}[k], \
g_{I}(\al^{*}_{I\mu}[k])g_{I}^{-1}=\al^{*}_{I\mu}[k]
\label{2.trivmap1}
\enr
We endowed the $"bc\bet\gm"$ fields in the formulas (\ref{2.vacc}),(\ref{2.nabla}) and (\ref{2.trivmap1}) by additional index $I$ because they differ for different cones $C_{I}$.
 
 As a consequence, $g_{I}$ defines the isomorphism between a pair of $N=2$ Virasoro superalgebras, where the second one acts on $M_{I}$ by the currents (\ref{2.btgmvir}) while the first one acts on $\mathbb{M}_{I}$ by the currents (\ref{2.btgmvir}) where the fields 
$\nabla _{\mu}(z)$ are taken instead of $a^{*}_{\mu}(z)$.

 Now the expression (\ref{2.twistellt}) follows from the corresponding $\check{C}$hech complex of the covering \cite{P2}. 
 
 One can use the isomorphisms (\ref{2.trivmap}) also to argue that the open string states on toric manifold with holomorphic line bundle can be described as line bundle twisted chiral de Rham complex.
To this end one should rewrite first the chiral de Rham complex in the logarithmic coordinates intensively used in \cite{B1}. It is given by the following expressions
\ber
a_{I\mu}(z)=\exp{[w^{*}_{I\mu}\cdot X]}(z),\ \al_{I\mu}(z)=w^{*}_{\mu}\cdot\psi\exp{[w^{*}_{I\mu}\cdot X]}(z),
\nmb
a^{*}_{I\mu}(z)=(e_{\mu}\cdot\d X^{*}-w^{*}_{I\mu}\cdot\psi e_{\mu}\cdot\psi^{*})
\exp{[-w^{*}_{I\mu}\cdot X]}(z), \
\al^{*}_{I\mu}(z)=e_{\mu}\cdot\psi^{*}\exp{[-w^{*}_{I\mu}\cdot X]}(z)
\label{2.btgmlog}
\enr
where $w^{*}_{I\mu}$ are the dual vectors to the basis of vectors $\left\{e_{\mu},\mu=0,...,\hat{I},...d\right\}$ generating the cone $C_{I}$:
\ber
<w^{*}_{I\mu},e_{\nu}>=\dlt_{\mu \nu}
\label{2.edual}
\enr
and $X_{i}(z), X^{*}_{i}(z)$, $i=1,2,...,d$ be the free bosonic fields and $\psi_{i}(z), \psi^{*}_{i}(z)$, $i=1,2,...,d$ be the
free fermionic fields:
\ber
X^{*}_{i}(z_{1})X_{j}(z_{2})=\ln(z_{12})\dlt_{i,j}+reg.,\nmb
\psi^{*}_{i}(z_{1})\psi_{j}(z_{2})=z_{12}^{-1}\dlt_{i,j}+reg,
\label{2.logope}
\enr
where $z_{12}=z_{1}-z_{2}$. It is clear that the bosons $X_{i}(z), X^{*}_{i}(z)$ correspond to the logarithmic (or polar) holomorphic and anti-holomorphic coordinates on the affine space $A_{I}$. It is easy to see due to (\ref{2.btgmlog}) that for each fixed positive value of $L[0]$-grading the corresponding subspace of the space of states generated from vacuum $|\Om_{I}>$ by the creation operators of $"bc\bet\gm"$ fields isomorphic to certain subspace
generated by the creation operators of the fields $X_{i}(z), X^{*}_{i}(z)$ and  $\psi_{i}(z), \psi^{*}_{i}(z)$. So the only difference appears in $L[0]=0$ grading because the general vertex
$\exp[p^{*}\cdot X(z)+p\cdot X^{*}(z)]$ contains also target space anti-holomorphic contributions when $p\neq 0$. Because of we consider only holomorphic Chan-Paton bundles this contribution vanishes.  Thus, it is natural to expect that the open string states on toric manifold with holomorphic line bundle can be described locally as a line bundle twisted chiral de Rham complex.
 
\vskip 10pt
\leftline{\it 2.3. Elliptic genus of line bundle twisted chiral de Rham complex on CY hypersurface.}  
\vskip 10pt

 Recall first what data CY hypersurface in toric variety is determined \cite{Batyr}. 
 
Let $\Lm_{1}$ and $\Lm^{*}_{1}$ be dual lattices of rank $d$. Denote by $\Lm$ and $\Lm^{*}$
two dual lattices such that $\Lm=\Lm_{1}\oplus \mathbb{Z}$ and $\Lm^{*}=\Lm^{*}_{1}\oplus \mathbb{Z}$. Let us denote by $deg$ the vector $(0,1)\in \Lm $ and by $deg^{*}$ we denote
the vector $(0,1)\in \Lm^{*}$. The two dual reflexive polytopes $\Delta \subset \Lm_{1}$ and $\Delta^{*}\in \Lm^{*}_{1}$ are given. The polytope $\Delta$ codes some toric variety $P_{\Delta}$ while the polytope $\Delta^{*}$ determine CY hypersurface in $P_{\Delta}$.
Let $K$ be a cone over $(\Delta,1)$ and $K^{*}$ be a cone over $(\Delta^{*},1)$. There is a complete fan in $\Lm_{1}$ whose one-dimensional cones are generated by all vertices of $\Delta$ and some additional points
in $\Delta$. This fan induces the decomposition $\Sgm$ of the cone $K$ into subcones such that each of the subcones includes $deg$. These are the toric data of the anticanonical bundle total space over $P_{\Delta}$. 

 The last ingredient is the toric divisor support function
$\om^{*}$ defined on $K$ which is a collection $\om^{*}_{I}$ of elements from $\Lm^{*}$ compatible with the restriction map for every intersection $C_{I}\cap C_{J}$ of maximal dimension cones from $\Sgm$ (see (\ref{2.supportf})). The toric divisor support function $\om^{*}$ determines a line bundle on anticanonical bundle total space. The line bundle on CY hypersurface is an induced bundle due to embedding defined by $\Delta^{*}$.

 The elliptic genus of line bundle twisted chiral de Rham complex on CY hypersurface is given by \cite{P2}
\ber
y^{-\frac{d-1}{2}}\sum_{p^{*}\in \Lm^{*}}t^{p^{*}-\om^{*}}
\sum_{C\subset\Sgm}(-1)^{codim C}\sum_{k\in C}
y^{-<deg^{*},k>+<deg,p^{*}>}q^{<p^{*},k>}G(y^{-1},q)^{d+1}
\label{2.Elltorcy}
\enr
This expression generalize the elliptic genus formula from Proposition 6.2 of \cite{B2}
for the case of line bundle twisted chiral de Rham complex on a hypersurface in toric variety
determined by the combinatorial data above. The proof of (\ref{2.Elltorcy}) goes similar
to the proof of Proposition 6.2 from \cite{B2}.

\vskip 10pt
\centerline{\bf 3. Line bundle twisted chiral de Rham complex elliptic genera calculations.}
\vskip 10pt

\vskip 10pt
\leftline{\it 3.1. Elliptic genus of line bundle twisted chiral de Rham complex on $\mathbb{P}^{1}$.}  
\vskip 10pt

 Let $e$ be the standard basis in $\mathbb{R}^{1}$. We fix the lattice
\ber
\Lm=\mathbb{Z}e
\label{3.P1latt}
\enr
and the dual lattice $\Lm^{*}$ generated by the basic vector $e^{*}$. 
The fan $\Sgm$ of $\mathbb{P}^{1}$ is the collection of 1-dimensional cones $C_{+}\in \Lm$, $C_{-}\in \Lm$ and 0-dimensional cone $C_{\bullet}=C_{+}\cap C_{-}$.
The $1-dim$ cones generating the fun $\Sgm$
are spanned by the vectors
\ber
C_{+}=Cone (e),
\
C_{-}=Cone (-e),
\label{3.1cones}
\enr

 The toric divisor support function defining the line bundle $O(N)$ on $\mathbb{P}^{1}$ is the collection $(\om^{*}_{+}, \om^{*}_{-})$ of elements from $\Lm^{*}$
\ber
\om^{*}_{+}=N_{+}e^{*},
\
\om^{*}_{-}=N_{-}e^{*}
\label{3.omP1}
\enr
where $N_{\pm}\in \mathbb{Z}$ and $N=N_{+}+N_{-}$. 
 
 According to (\ref{2.twistellt}) we find
\ber
Ell_{\om^{*}}(\mathbb{P}^{1},t,y,q)=y^{-1\ov 2}\sum_{n\in \mathbb{Z}}
(\frac{t^{n-N_{+}}}{1-yq^{n}}G(y,q)+\frac{t^{-n-N_{-}}}{1-yq^{n}}G(y,q)-G(y,q))
\label{3.P1ell}
\enr
One can rewrite this expression in terms of theta functions using the identity (\ref{2.Ident}) and  riding of the contribution from $C_{\bullet}$ \cite{B2}, \cite{P2}
\ber
Ell_{\om^{*}}(\mathbb{P}^{1},t,y,q)=t^{-N_{+}}\frac{\Tt(ty^{-1},q)}
{\Tt(t,q)}+t^{N-N_{+}}\frac{\Tt(t^{-1}y^{-1},q)}
{\Tt(t^{-1},q)}
\label{3.ElltorP1}
\enr
where
\ber
\Tt(u,q)=q^{1/8}\prod_{n=0}(1-u^{-1}q^{n+1})(1-uq^{n})(1-q^{n+1})=
q^{1/8}\sum_{n\in Z}(-1)^{n}q^{(n^{2}-n)/2}u^{-n},
\label{2.Theta}
\enr

 By the l'Hopital rule we find
\ber
Ell_{N}(\mathbb{P}^{1},y,q)\equiv lim_{t \rightarrow 1}Ell_{\om^{*}}(\mathbb{P}^{1},t,y,q)=
Ny\et(q)^{-3}\Tt(y,q)+Ell(\mathbb{P}^{1},y,q),
\label{3.P1Ell}
\enr
where
\ber
Ell(\mathbb{P}^{1},y,q)=y\et(q)^{-3}(\Tt(y,q)+2y\frac{\d \Tt(y,q)}{\d y}),
\nmb
\et(q)=q^{1\ov 24}\prod_{n=1}(1-q^{n})
\label{3.P1Ell1}
\enr

 One can give the following interpretation of the expression (\ref{3.P1Ell}). The last term is just the elliptic genus of $\mathbb{P}^{1}$ coming from $D2$-brane wrapping $\mathbb{P}^{1}$. The first term gives the contribution due to nontrivial line bundle $O(N)$ is defined on $\mathbb{P}^{1}$. It coincides with the open string oscillators contribution coming from $N$ $D0$-branes on $\mathbb{P}^{1}$. Indead, we see that only oscillator string exitations contribute and there is no open string zero modes contribution. Thus we have Dirishlet boundary conditions. It allows us to interprate the cohomology of chiral de Rham complex twisted by $O(N)$-bundle as open string states of the bound state of $N$ toric invarianr $D0$-branes and one $D2$- brane on $\mathbb{P}^{1}$.

\vskip 10pt
\leftline{\it 3.2. Elliptic genus of line bundle twisted chiral de Rham complex on two dimensional}
\leftline{\it toric manifold.}  
\vskip 10pt

 We calculate first the elliptic genus of line bundle twisted chiral de Rham complex on $\mathbb{P}^{2}$.
 
 Let $e_{1},e_{2}$ be the standard basis in $\mathbb{R}^{2}$. We fix the lattice
\ber
\Lm=\mathbb{Z}e_{1}\oplus\mathbb{Z}e_{2}\subset\mathbb{R}^{2}
\label{3.P2latt}
\enr
and its dual lattice $\Lm^{*}$. The $2-dim$ cones generating the fun $\Sgm$
are spanned by the vectors
\ber
C_{01}=Cone (e_{0}=-e_{1}-e_{2},e_{1}),
\nmb
C_{02}=Cone (e_{0}=-e_{1}-e_{2},e_{2}),
\nmb
C_{12}=Cone (e_{1},e_{2})
\label{3.2con}
\enr
The $1-dim$ cones are
\ber
C_{2}=C_{12}\cap C_{02}=Cone (e_{2}),
\nmb
C_{1}=C_{12}\cap C_{01}=Cone (e_{1}),
\nmb
C_{0}=C_{01}\cap C_{02}=Cone (e_{0})
\label{3.1con}
\enr
The $0$-dimensional cone is $C_{\bullet}=C_{0}\cap C_{1}\cap C_{2}$.

 The toric divisor support function $\om^{*}$ of the line bundle $O(N)$ is determined by its values on the vectors generating $1$-dimensional cones
\ber
\om^{*}(e_{1})=N_{1},
\
\om^{*}(e_{2})=N_{2},
\
\om^{*}(e_{0})=N_{0},
\nmb
N=N_{0}+N_{1}+N_{2}
\label{3.omP2}
\enr

 According to (\ref{2.twistellt}) we obtain 
\ber
Ell_{N}(\mathbb{P}^{2},t_{1},t_{2},y^{-1},q)=
t_{1}^{-N_{1}}t_{2}^{-N_{2}}\frac{\Tt(t_{1}y^{-1},q)\Tt(t_{2}y^{-1},q)}
{\Tt(t_{1},q)\Tt(t_{2},q)}+
\nmb
t_{1}^{-N_{1}}t_{2}^{N_{0}+N_{1}}\frac{\Tt(t_{2}^{-1}y^{-1},q)
\Tt(t_{1}t_{2}^{-1}y^{-1},q)}
{\Tt(t_{2}^{-1},q)\Tt(t_{1}t_{2}^{-1},q)}+
t_{1}^{N_{0}+N_{2}}t_{2}^{-N_{2}}\frac{\Tt(t_{1}^{-1}y^{-1},q)
\Tt(t_{1}^{-1}t_{2}y^{-1},q)}
{\Tt(t_{1}^{-1},q)\Tt(t_{1}^{-1}t_{2},q)}
\label{3.Elltor2}
\enr
By the l'Hopital rule we find
\ber
Ell_{N}(\mathbb{P}^{2},y,q)\equiv lim_{t_{1},t_{2} \rightarrow 1}Ell_{\om^{*}}(\mathbb{P}^{2},t_{1},t_{2},y,q)=
\nmb
\frac{N^{2}}{2}(y\et(q)^{-3}\Tt(y,q))^{2}+\frac{3N}{2}(y\et(q)^{-3}\Tt(y,q))Ell(\mathbb{P}^{1},y,q)+Ell(\mathbb{P}^{2},y,q)
\label{3.EllP2ND}
\enr
where 
\ber
Ell(\mathbb{P}^{2},y,q)=(\frac{9}{8}-3q\et(q)^{-1}\frac{\d \et(q)}{\d q})(y\et(q)^{-3}\Tt(y,q))^{2}+
\nmb
y^{3}\et(q)^{-6}(6\Tt(y,q)\frac{\d \Tt(y,q)}{\d y}+
3y(\frac{\d \Tt(y,q)}{\d y})^{2}+\frac{3}{2}y\Tt(y,q)\frac{\d^{2} \Tt(y,q)}{\d y^{2}})
\label{3.EllP2}
\enr
is the Elliptic genus of $\mathbb{P}^{2}$.

 We give the following $D$-brane interpretation of the expression above. Similar to the line bundle twisted chiral de Rham complex on $\mathbb{P}^{1}$, the first contribution comes from the open string states of $N^{2}$ $D0$-branes. The second one comes from the open string states of $D2$-brane which is a divisor linearly equivalent to $N$ multiple of the hyperplane $\mathbb{P}^{1}\subset \mathbb{P}^{2}$, so the factor $y\et(q)^{-3}\Tt(y,q)$
gives the open string oscillator contributions in the transverse direction to the $D2$-brane. Because of there is no open string zero modes contribution we have Dirishlet boundary conditions in the transverse direction. The last term comes from $D4$-brane wrapping $\mathbb{P}^{2}$. Thus the cohomology of chiral de Rham complex twisted by $O(N)$ bundle describes the open string states of the bound state of $D0$-$D2$-$D4$-branes. 

 An obvious generalization of (\ref{3.EllP2ND}) for the case of chiral de Rham complex twisted by line bundle $O(D)$ on two dimensional complete smooth toric variety
$\mathbb{P}_{\Delta}$ is given by
\ber
Ell_{D}(\mathbb{P}_{\Delta},y^{-1},q)=
\nmb
{D^{2}\ov 2}(y\eta(q)^{-3}\Tt (y,q))^{2}-{K\cdot D\ov 2}(y\et(q)^{-3}\Tt(y,q))Ell(\mathbb{P}^{1},y^{-1},q)+Ell(\mathbb{P}_{\Delta},y^{-1},q)
\label{3.PDeltaEll}
\enr
where $D$ is the divisor of a line bundle $O(D)$ and $K$ is canonical bundle divisor of $\mathbb{P}_{\Delta}$. 

 In this more general situation we see again the open string contributions from $D^{2}$  $D0$-branes, one $D2$-brane wrapping the divisor $D$ and one $D4$-brane wrapping $\mathbb{P}_{\Delta}$. It is interesting to note here a similarity of the expression (\ref{3.PDeltaEll}) to the expression for the Euler characteristic of a line bundle which is given by the Riemann-Roch theorem for surfaces \cite{GrH}. Moreover, in the limit $q\rightarrow 0$ and $y\rightarrow 1$ the formula (\ref{3.PDeltaEll}) reproduces it exactly. 
 
\vskip 10pt
\leftline{\it 3.3. Elliptic genus of line bundle twisted chiral de Rham complex on K3.}  
\vskip 10pt

 In this subsection the elliptic genus of
line bundle twisted chiral de Rham complex on K3 hypersurface in $\mathbb{P}^{3}$ is calculated. We begin by specifying the toric data from Subsection 2.3.

 Let $\lbrc e_{1},...,e_{4}\rbrc$ be the standard basis $\mathbb{R}^{4}$. Let $\Lm$ be the lattice
in $\mathbb{R}^{4}$ generated by the vectors $e_{0}={1\ov 4}(e_{1}+...e_{4})$, $e_{1}$,...,$e_{4}$ and $\Lm^{*}$ be the dual lattice. $deg=e_{0}$ and $deg^{*}=e^{1}+...+e^{4}$, where $ \lbrc e^{1},...,e^{4}\rbrc$ is the dual basis to $\lbrc e_{1},...,e_{4}\rbrc$. Hence $\Lm=\mathbb{Z}deg\oplus\Lm_{1}$ and $\Lm^{*}=\mathbb{Z}deg^{*}\oplus\Lm_{1}$.
The vertices of the polytope $\Delta\subset \Lm_{1}$ are given by the vectors $e_{0},...,e_{4}$, so that $P_{\Delta}=\mathbb{P}^{3}$. We fix also the dual reflexive polytope $\Delta^{*}\subset \Lm^{*}_{1}$ with the only internal point $deg^{*}$. According to subsection 2.3. we determine the cones $K$, $K^{*}$ and fan $\Sgm$. 

  The toric divisor support function $\om^{*}$ of the line bundle $O(N)$ induced on $\mathbb{P}^{3}$ is determined by its values on the basic vectors generating $1$-dimensional cones from $\Sgm$
\ber
\om^{*}(e_{i})=N_{i},
\
i=1,...,4,
\
\om^{*}(e_{0})=N_{0},
\nmb
N=N_{1}+...+N_{4}-4N_{0}
\label{3.omK3}
\enr

 Specifying (\ref{2.Elltorcy}) to the case at hand and riding off the
positive codimension cones contributions (see \cite{B2}, \cite{P2}) we obtain 
\ber
Ell_{\om^{*}}(K3,t_{1},...,t_{4},y,q)=y^{2}\sum_{I=1}^{4}(\prod_{i=1}^{4}t_{i}^{-<\om_{I}^{*},e_{i}>})
\frac{\Tt(t_{I}^{4})}{\Tt(t_{I}^{4}y)}\prod_{J\neq I}\frac{\Tt(t_{I}^{-1}t_{J}y^{-1})}{\Tt(t_{I}^{-1}t_{J})}=
\nmb
y^{2}(t_{1}^{N}t_{2}^{-N_{2}}...t_{4}^{-N_{4}}
\frac{\Tt(t_{1}^{4})}{\Tt(t_{1}^{4}y)}\frac{\Tt(t_{1}^{-1}t_{2}y^{-1})}{\Tt(t_{1}^{-1}t_{2})}
\frac{\Tt(t_{1}^{-1}t_{3}y^{-1})}{\Tt(t_{1}^{-1}t_{3})}
\frac{\Tt(t_{1}^{-1}t_{4}y^{-1})}{\Tt(t_{1}^{-1}t_{4})}+
\nmb
t_{1}^{-N_{1}}t_{2}^{N}t_{3}^{-N_{3}}t_{4}^{-N_{4}}
\frac{\Tt(t_{2}^{4})}{\Tt(t_{2}^{4}y)}\frac{\Tt(t_{2}^{-1}t_{1}y^{-1})}{\Tt(t_{2}^{-1}t_{1})}
\frac{\Tt(t_{2}^{-1}t_{3}y^{-1})}{\Tt(t_{2}^{-1}t_{3})}
\frac{\Tt(t_{2}^{-1}t_{4}y^{-1})}{\Tt(t_{2}^{-1}t_{4})}+
\nmb
t_{1}^{-N_{1}}t_{2}^{-N_{2}}t_{3}^{N}t_{4}^{-N_{4}}
\frac{\Tt(t_{3}^{4})}{\Tt(t_{3}^{4}y)}\frac{\Tt(t_{3}^{-1}t_{1}y^{-1})}{\Tt(t_{3}^{-1}t_{1})}
\frac{\Tt(t_{3}^{-1}t_{2}y^{-1})}{\Tt(t_{3}^{-1}t_{2})}
\frac{\Tt(t_{3}^{-1}t_{4}y^{-1})}{\Tt(t_{3}^{-1}t_{4})}+
\nmb
t_{1}^{-N_{1}}t_{2}^{-N_{2}}t_{3}^{-N_{3}}t_{4}^{N}
\frac{\Tt(t_{4}^{4})}{\Tt(t_{4}^{4}y)}\frac{\Tt(t_{4}^{-1}t_{1}y^{-1})}{\Tt(t_{4}^{-1}t_{1})}
\frac{\Tt(t_{4}^{-1}t_{2}y^{-1})}{\Tt(t_{4}^{-1}t_{2})}
\frac{\Tt(t_{4}^{-1}t_{3}y^{-1})}{\Tt(t_{4}^{-1}t_{3})})
\label{3.EllK3tor}
\enr
 
 The limit $(t_{1},...,t_{4})\rightarrow (1,1,1,1)$ is given by
\ber
Ell_{N}(K3,y,q)=2N^{2}(y\et(q)^{-3}\Tt_{1,1}(y,q))^{2}+Ell(K3,y,q)
\label{3.EllK3N}
\enr
where 
\ber
Ell(K3,y,q)=
48y^{2}q\frac{\d \log{\et(q)}}{\d q}(\et(q)^{-3}\Tt(y))^{2}
\nmb
-24y^{3}\et(q)^{-6}(\Tt(y)\frac{\d \Tt(y)}{\d y}-y(\frac{\d \Tt(y)}{\d y})^{2}+
y\Tt(y)\frac{\d^{2} \Tt(y)}{\d y^{2}})
\label{3.EllK3}
\enr
is the elliptic genus of $K3$.

 One can see the similarity between the results (\ref{3.PDeltaEll}) and (\ref{3.EllK3N}).
There is a difference however because of the canonical bundle of $K3$ is trivial so that we have the open string contributions coming from $4N^{2}$ $D0$-branes and one $D4$-brane on $K3$.


\vskip 10pt
\centerline{\bf 4. Elliptic genera for more general twisting sheaves.}
\vskip 10pt

 Here we generalize the results of the preceding section to include more general examples of twisting sheaves. The explicit calculations are given in three examples. In the first example the twisting sheaf localized on a curve in $\mathbb{P}^{2}$. In the second example the chiral de Rham complex is twisted by a sheaf localized on points in $\mathbb{P}^{2}$. In the third example we twist the chiral de Rham complex by $SU(N)$ vector bundle with the instanton number $k$. 
 
 The idea behind the calculation is simple. To construct more general twisting sheaf we use a sequence of locally free sheaves. Then we assume that the sequence can be extended to a sequence of twisted $MSV(X)$ sheaves and hence, the elliptic genus we are interested in is given by (alternating) sum of elliptic genera of the twisted $MSV(X)$ sheaves calculated in the section above. Though, we are not giving the proof, the structure of the formulas (\ref{3.P1Ell}), (\ref{3.EllP2ND}), (\ref{3.PDeltaEll}) obviously confirms the assumption if we read them as a $D0$ and $D2$- branes or $D2$-brane and $D4$-brane binded together to form the bound state  due to tachyon condensation \cite{Sen} (see also \cite{Asp}, \cite{Sh} and references therein) so that the open string states are given by the cohomology of line bundle twisted chiral de Rham complex.

\vskip 10pt
\leftline{\it 4.1. Elliptic genus of chiral de Rham complex twisted by a sheaf localized on a curve.}  
\vskip 10pt

 Let us consider the exact sequence determined on $\mathbb{P}^{2}$
\ber
0\rightarrow O \stackrel{T}{\rightarrow} O(N)\rightarrow O_{D}(N)\rightarrow 0
\label{4.seq}
\enr
The divisor $D$ of the line bundle $O(N)$ is fixed by the equation $T=0$ and $O_{D}(N)$ is a sheaf localized on $D$. To calculate the elliptic genus of chiral de Rham complex twisted by the sheaf $O_{D}(N)$ we assume that the exact sequence (\ref{4.seq}) can be extended to the exact sequence of twisted $MSV(\mathbb{P}^{2})$ sheaves
\ber
0\rightarrow MSV(\mathbb{P}^{2})\stackrel{\hat{T}}{\rightarrow} O(N)\otimes MSV(\mathbb{P}^{2})
\rightarrow O_{D}(N)\otimes MSV(\mathbb{P}^{2})\rightarrow 0
\label{4.MSVseq}
\enr
where $\hat{T}$ is some vertex operator.
Then the elliptic genus $Ell^{*}_{N}(\mathbb{P}^{2},y,q)$ of chiral de Rham complex twisted by $O_{D}(N)$ is given by
\ber
Ell^{*}_{N}(\mathbb{P}^{2},y,q)=
Ell_{N}(\mathbb{P}^{2},y,q)-Ell(\mathbb{P}^{2},y,q)
\label{4.EllD}
\enr
Using (\ref{3.EllP2ND}) we find
\ber
Ell^{*}_{N}(\mathbb{P}^{2},y,q)=
\frac{N^{2}}{2}(y\et(q)^{-3}\Tt(y,q))^{2}+\frac{3N}{2}(y\et(q)^{-3}\Tt(y,q))Ell(\mathbb{P}^{1},y,q)
\label{4.EllD1}
\enr
Thus we see the open string contributions coming from bound state of $N^{2}$ $D0$-branes and one $D2$-brane which is linearly equivalent to $N$ multiple of the hyperplane $\mathbb{P}^{1}$ in $\mathbb{P}^{2}$. 

 The result is clearly generalized for a more general two dimensional toric manifold.
 
\vskip 10pt
\leftline{\it 4.2. Elliptic genus of chiral de Rham complex twisted by a sheaf localized on points.}  
\vskip 10pt

 Let us consider for example the sheaf 
\ber
J=\mathbb{C}[x,y]/I,
\label{4.sheafJ}
\enr
where the ideal sheaf $I$ is determined in local coordinates $x_{1},x_{2}$ on $\mathbb{P}^{2}$ by
\ber
I=\lbrc x_{2}^{4}p_{1}(x_{1},x_{2})+x_{1}x_{2}^{3}p_{2}(x_{1},x_{2})+x_{1}^{2}x_{2}p_{3}(x_{1},x_{2})+x_{1}^{4}p_{4}(x_{1},x_{2})\rbrc
\label{4.Ideal}
\enr
where $p_{i}(x_{1},x_{2})$ are the polynomials. 
The sheaf $J$ can also be represented as a cohomology of the following complex
\ber
0\rightarrow 
\left(\begin{array}{c}
      O(-5) \\
      O(-5) \\
      O(-5)
      \end{array}\right)
\stackrel{d_{1}}{\rightarrow}
\left(\begin{array}{c}
      O(-4) \\
      O(-3) \\
      O(-4) \\
      O(-4)
      \end{array}\right)
\stackrel{d_{2}}{\rightarrow}O\rightarrow 0
\label{4.Kozul}
\enr
where the differential $d_{1}$ is given by the matrix
\ber
d_{1}=\left(\begin{array}{ccc}
      x_{2}     &0    &0 \\
      -x_{1}^{2}&x_{2}^{2}&0 \\
      0     &-x_{1}   &x_{2} \\
      0     &0    &-x_{1}
      \end{array}\right)
\label{4.d1}
\enr
and the differential $d_{2}$ is given by the vector
\ber
d_{2}=(x_{1}^{4},x_{1}^{2}x_{2},x_{1}x_{2}^{3},x_{2}^{4})
\label{4.d2}
\enr
Notice that the only nonzero Chern class of the sheaf $J$ is the second one
$ch_{2}(F)=9$.

 Due to the assumption that complex (\ref{4.Kozul}) can be generalized to the complex
of line bundle twisted $MSV(\mathbb{P}^{2})$ sheaves, the elliptic genus of the chiral de Rham complex twisted by $J$, is given by
\ber
Ell_{J}(\mathbb{P}^{2},y,q)=
Ell(\mathbb{P}^{2},y,q)-3Ell_{(-4)}(\mathbb{P}^{2},y,q)-
\nmb
Ell_{(-3)}(\mathbb{P}^{2},y,q)+3Ell_{(-5)}(\mathbb{P}^{2},y,q)=
9(y\et(q)^{-3}\Tt(y,q))^{2}  
\label{4.Ellpoint}
\enr
where we have used (\ref{3.EllP2ND}).

 The $D$-brane interpretation of this result is obvious: the elliptic genus is given by the open string oscillator contributions from the bound state of 9 $D0$-branes.
 
 Clearly this result can be generalized for more general toric manifold.

\vskip 10pt
\centerline{\it 4.3. Elliptic genus of chiral de Rham complex on $\mathbb{P}^{2}$ twisted by}
\centerline{\it $k$-instanton $SL(N)$ vector bundle.}
\vskip 10pt

 Due to Donaldson \cite{Don} the set of sections of $k$-instanton $SL(N)$ vector bundle on $\mathbb{P}^{2}$ can be given 
by the cohomology of the complex
\ber
O(-1)^{\oplus k}\stackrel{d_{1}}{\rightarrow} O^{\oplus (2k+N)}\stackrel{d_{2}}{\rightarrow} O(1)^{\oplus k}
\label{F.monad}
\enr
where
\ber
d_{1}=A_{0}X_{0}+A_{1}X_{1}+A_{2}X_{2},
\nmb
d_{2}=B_{0}X_{0}+B_{1}X_{1}+B_{2}X_{2}
\label{F.maps}
\enr
where $X_{0},X_{1},X_{2}$ are the 
homogeneous coordinates on $\mathbb{P}^{2}$ and $A_{i}$, $B_{i}$, $i=0,1,2$ are the constant matrices satisfying $d_{1}d_{2}=0$. 

 Assuming that complex (\ref{F.monad}) can be generalized to the complex
of line bundle twisted $MSV(\mathbb{P}^{2})$ sheaves one can
write out the elliptic genus of the chiral de Rham complex twisted by $k$-instanton $SL(N)$ vector bundle as the alternating sum
\ber
Ell_{(SL(N),k)}(\mathbb{P}^{2},y,q)=
\nmb
kEll_{(-1)}(\mathbb{P}^{2},y,q)-
(2k+N)Ell(\mathbb{P}^{2},y,q)+kEll_{(1)}(\mathbb{P}^{2},y,q)
\label{F.Ell}
\enr
Using (\ref{3.EllP2ND}) we obtain
\ber
Ell_{(SL(N),k)}(\mathbb{P}^{2},y,q)=k(y\et(q)^{-3}\Tt(y,q))^{2}-
NEll(\mathbb{P}^{2},y,q)
\label{F.EllSLN}
\enr
Thus we have a bound state of $k$ $D0$-branes and $N$ $D4$-antibranes wrapping $\mathbb{P}^{2}$
which is in agreement with Witten's conjecture \cite{W}.

 In conclusion of this section some comment on tachyon condensation \cite{Sen}, \cite{BMOz}
is in order. In the topological open string B model the complex of sheaves correspond to a set of topological branes and antibranes. Taking the cohomology of the complex models brane-antibrane annihilation due to tachyon condensation (see \cite{Asp}, \cite{Sh} and references therein). In this approach the only open string states one can observe are the ground states. The infinite tower of open string states erected over the ground ones are canceled because of topological reduction. On the contrary, the results of this section show that one can reproduce the infinite tower of open string states taking the sequences
of twisted $MSV$ sheaves instead of usual ones and taking corresponding vertex operators modeling the tachyons. In other words, making such substitution we reproduce brane-antibrane annihilation for the physical $D$-branes, at least in the large radius limit of the toric manifold.

\vskip 10pt
\centerline{\bf 5. Conclusion.}
\vskip 10pt

 In this note we calculated elliptic genus in various examples of twisted chiral de Rham complex on two dimensional toric compact manifolds and Calabi-Yau hypersurface in compact smooth toric manifold. In all cases considered, we found an infinite tower of open string oscillator contributions coming from the corresponding bound states of $D$-branes and 
identified directly the open string boundary conditions to the characteristic classes of Chan-Paton bundles. Our results confirm the conjecture of \cite{P2} that twisted chiral de Rham complex describes an infinite tower of states in the open string sector of bound state of $D$-branes in the large radius limit of boundary sigma model on toric manifold. 

 One of the obvious question remained open is to explain in more details the geometric meaning of the elliptic genus expressions like (\ref{3.EllP2ND}), (\ref{3.PDeltaEll}), (\ref{3.EllK3N}). It would be interesting in particular to interprate these expressions as string generalization of Hirzebruch-Riemann-Roch index formula. Similar question in the context of topological boundary B model has been discussed in the works of
\cite{Cald1} and \cite{Cald2}, see also \cite{Sh}. 

There is also a technical problem of elliptic genus calculation for higher dimensional toric manifolds. It would be interesting in particular to calculate the elliptic genus of twisted chiral de Rham complex on a toric Calabi-Yau three-fold. The proof of the assumption the calculations of section 4 were based on is another open question. 

\vskip 20pt
\leftline{\bf Acknowledgements}
\vskip 10pt

I thank Boris Feigin and Mikhail Bershtein for helpful discussions.
The study was supported, in part, by the Ministry of Education and Science of
Russian Federation under the contracts No.8410 and No.8528 and by Russian Foundation for Basic Research
under the grant No.13-01-90614.

\end{document}